\documentclass[11pt]{article}
\usepackage{moriond,epsfig}

\def\Journal#1#2#3#4{{#1} {\bf #2}, #3 (#4)}

\def\be{\begin{equation}}
\def\ee{\end{equation}}
\def\bea{\begin{eqnarray}}
\def\eea{\end{eqnarray}}

\begin{document}
\vspace*{4cm}
\title{STATUS REPORT ON THE NEMO 3 EXPERIMENT}

\author{ANNE-ISABELLE ETIENVRE}

\address{Laboratoire de l'Acc\'el\'erateur Lin\'eaire, IN2P3-CNRS et Universit\'e Paris-Sud, Orsay, France}

\maketitle\abstracts{
The NEMO 3 detector, now operating in the Fr\'ejus Underground Laboratory, is devoted to search for neutrinoless double beta decay~; the expected sensitivity for the effective neutrino mass ($< m_{\nu} >$) is on the order of 0.1 eV. The performances of the tracking detector are presented. The first and very preliminary results concerning the background and the double beta signal are given.}

\section{Introduction}
\hspace{0.5cm}The recent discovery of neutrino oscillation is a proof that the neutrino is a massive particle. However, another fundamental question remains, concerning the nature of massive neutrinos~: are they Majorana particles ($\nu\, = \, \bar{\nu}$) or Dirac particles ($\nu\, \neq \, \bar{\nu}$)? Neutrinoless double beta decay ($\beta\beta0\nu$), which consists in the decay of an (A,Z) nucleus to an (A, Z+2) nucleus by the simultaneous emission of two electrons but without the emission of neutrinos, is a process forbidden by the Minimal Standard Model~; the observation of $\beta\beta0\nu$ would be a signature of lepton number non-conservation. Moreover, under the assumption that the mechanism responsible for $\beta\beta0\nu$ lies on the exchange of a massive neutrino, this observation would prove that neutrinos are Majorana particles.

%
%
\section{The NEMO 3 experiment}

\subsection{Description}

\begin{figure}[h]  
         \centerline{\epsfxsize=7cm \epsfbox{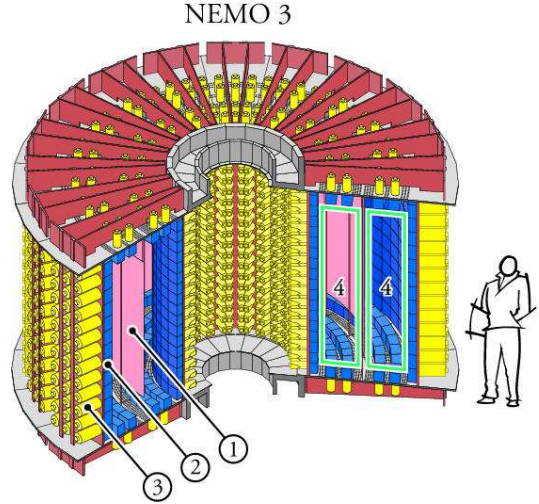}} 
         \caption{\label{nemo} \textit{Schematic view of the NEMO3 detector~: \,(1) Source foil  \, (2)  \, Calorimeter made of 1940 plastic scintillators coupled to  (3) low activity photomultipliers \, (4) Tracking volume with 6180 drift cells operating in Geiger mode }}
\end{figure}

\hspace{0.5cm}The NEMO 3 detector \cite{nemo} is cylindrical and divided into 20 equal sectors (Fig.\ref{nemo}) . A thin (40-60 mg/$cm^2$) cylindrical enriched source foil of $\beta\beta$ emitters has been constructed from either a metal film or powder bound by an organic glue to mylar strips. The detector houses 10 kg of these isotopes. The main nucleus studied in NEMO 3 is $^{100}Mo$ ~: 6.9 kg of enriched (the enrichment factor is equal to 97$\%$) $^{100}Mo$ have been introduced in NEMO 3. \\

	The source hangs between two concentric cylindrical tracking volumes consisting of 6180 open octogonal drift cells operating in geiger mode, which provide a three dimensionnal reconstruction of the tracks. The tracking volume is surrounded by the calorimeter, made of 1940 blocks of plastic scintillators coupled to very low radioactivity photomultipliers~; the energy and time resolutions are roughly equal to 60 keV and 250 ps, respectively, at 1 MeV.  A laser calibratoin system permits daily checks on the stability of the energy and time calibration parameters.\\
	The charge recognition is assured by a vertical magnetic field of 30 gauss generated by  a solenoid that surrounds  the detector. A 20 cm external shield made of low radioactivity iron reduces the $\gamma$ ray flux~; 30 cm of borated water suppresses the contribution of fast neutrons.\\
	All the materials used  in the detector have been selected for their high radiopurity by $\gamma$-ray spectroscopy using HP Ge detectors.\\
	The NEMO 3 detector is able to recognize electrons, positrons, $\alpha$-particles and photons.
	
\subsection{Expected backgrounds in the $\beta\beta0\nu$ analysis}

The background events which can mimic $\beta\beta0\nu$ events (i.e. two electrons with no additional photon(s)) have three main contributions (the contribution of cosmic rays is negligible)~: neutrons, natural radioactivity ($^{208}Tl$, $^{214}Bi$), and the end-point region of the $\beta\beta2\nu$ spectrum. The requirements of the NEMO 3 experiment concerning the contribution of $^{208}Tl$, $^{214}Bi$ and neutrons are very strict \cite{neutron}~: the maximal radioactivity of the source foils allowed is equal to 20 $\mu$Bq/kg for the $^{208}$Tl and 300 $\mu$Bq/kg for the $^{214}$Bi. After 5 years of acquisition with 7 kg of $^{100}Mo$, the expected events due to the background, in the energy window (2.8 to 3.2 MeV) centered on the end-point of the isotope studied (named $Q_{\beta\beta}$)  are the following~: 1 event due to $^{208}Tl$ pollution, 1 event due to $^{214}Bi$ pollution and 0.03 event due to the residual neutrons. The purity of the foils in NEMO 3 makes the contribution of the $\beta\beta2\nu$ process the dominant background.\\

NEMO 3 is able to reconstruct its own background~: this will be illustrated later in this paper through the example of the source pollution in $^{208}$Tl.

\subsection{Running of the experiment}

In june 2002, the 20 sectors of NEMO 3, its coils and the iron shied were installed~:  the detector began to take its first test runs, which gave us the possibility to perform several preliminary analysis concerning $\beta\beta0\nu$, $\beta\beta2\nu$, and the different kinds of background, since the time and energy calibration of the detector had been done at the beginning of this running. In december 2002, we stopped the experiment and installed the neutron shield, and, since the $14^{th}$ of february, NEMO 3 is taking data in stable conditions.

\section{Performances of the tracking detector}

\subsection{Generation of high-energy electrons crossing}

An Am/Be neutron source, situated on the bottom of the detector,  emits fast neutrons, thermalized in the plastic of the scintillators~; then, a radiative capture of these thermalized neutrons, in the copper present inside the detector produces a $\gamma$ whose energy can go up to 8 MeV. The Compton electron created by this $\gamma$ can cross all the detector, from one wall of scintillators to the opposited one. Using crossing electron with an energy higher than 4.5 MeV, we can study properly the tracking reconstruction, since the effects of multiple scattering become negligible.\\

Using these selected electrons, we could determine the law between the drift time and the drift distance, inside a geiger cell, necessary for the transversal reconstruction of the tracks~: the drift distance is, excepted for very short ($\leq$ 200 ns) or very long ($\geq$ 1500 ns) drift times, roughly proportional to the square root of the drift time, which is logical since the electrostatic field inside a geiger cell is inversely proportionnal to the drift distance. \\

Looking at the distribution of the residue, defined as the difference between the drift distance and the transversal reconstructed distance, inside a geiger cell, we could determine the average transversal and longitudinal resolutions for a drift cell, equal to 0.4 mm and 0.8 cm, respectively.\\ 

The charge recognition, assured by the existence of a magnetic field, had to be checked~; therefore, we used the same sample of electron crossing events, and constrained the first part of the track - from the first wall of PM to the source foil -  to be reconstructed as an electron. Therefore, the study of the second part of the track - from the source foil to the opposite wall - gives us the probability to confuse an electron and a positron, equal to 3 $\%$ at 1 MeV.

\subsection{Spatial resolution on the vertex}

In each sector of the detector, there is a copper tube at the radius of the foil, which runs vertically for the height of the detector~; during energy calibration runs, each tube has inserted into it three $^{207}Bi$ sources of 5 nCi, for a  total of sixty sources. As the position of these sources are very well known, the study of the two conversion electrons (0.5 and 1 MeV) emitted by these sources gives us access to the determination of the spatial resolution on the vertex of the tracks. In the 1-electron channel, the transversal and longitudinal resolutions are equal to 0.2 cm and 0.8 cm, respectively, at 1 MeV. In the 2-electron channel, useful since the signal searched by NEMO 3 is precisely made of two electrons, they are equal to 0.6 cm and 1.0 cm, respectively.

\section{Radiopurity in $^{208}Tl$ of the sources of $^{100}Mo$}

\subsection{Principle}
Since a $^{208}Tl$ nucleus emits, in 100$\%$ of its desintegrations, a $\gamma$-ray with the highest energy of the natural radioactivity (2.6 MeV), close to the endpoint of the $\beta\beta$ process, situated around 3 MeV for the nuclei studied in NEMO 3, this nucleus is the most dangerous background for $\beta\beta0\nu$ study. Therefore, the pollution in $^{208}Tl$ of the foils has to be very well known. The principle of the study lies on the exploitation of the $e^-n\gamma$ ($1\, \leq\, n \, \leq \, 3$) channels, with appropriated cuts on the energy of the different particles, but also on the time of flight of them (through a $\chi^2$ analysis), in order to select events emitted in the source foil.\
The cuts applied on the total energy of the photon(s), on the energy of the electron, on the $\chi^2_{internal}$ that characterizes temporally events emitted in the foil sources, the efficiencies related to the three channels, are shown in Table \ref{resul_tl208}. We can consider that the possible background contribution to this analysis, which consists in external or internal $^{214}Bi$ and external $^{208}Tl$, is negligible with these cuts.

\begin{table}[htbp]
\begin{center}
\begin{tabular}{|l|c|c|c|c|c|c||}
\hline \hline
Channel  & Total energy & Energy of  & $\chi^2_{internal}$ & Efficiency ($\%$) \\
 & of the photons (MeV) &  the electron & &\\
\hline \hline
$e^-\gamma$  & $\geq$ 2.3 &$0.5\, \leq\, E_{e^-}\, \leq \, 1.3$ & $\leq$ 6.7&0.294 $\pm$ 0.005 \\
 \hline
$e^-2\gamma$  & $\geq$ 2.3&$0.5\, \leq\, E_{e^-}\, \leq \, 1.3$  &$\leq$ 13.8 &0.369 $\pm$ 0.006 \\
 \hline
$e^-3\gamma  $&  $\geq$ 2.6 & $0.5\, \leq\, E_{e^-}\, \leq \, 1.3$ & $\leq$ 16.3&0.114 $\pm$ 0.003 \\
 \hline

\hline

\end{tabular}
\end{center}
\caption{Characteristics of the differents channels studied for the determination of the $^{208}Tl$ pollution inside the $^{100}Mo$ source~: cuts applied and efficiency (with statistical error), determined using simulated events}
\label{resul_tl208}
\end{table} 	 

\subsection{Preliminary results}

Using 900 hours of data, taken in unstable conditions, we get the upper limits on the $^{208}Tl$ pollution of the $^{100}Mo$ sources shown in Table \ref{resul_tl208_2}~; if we combine the three channels ($e^-\gamma$, $e^-2\gamma$, $e^-3\gamma$), considering that the contribution of the other backgrounds (external or internal $^{214}Bi$ and external $^{208}Tl$) to the search for an internal $^{208}Tl$ signal is negligible, the upper limit on the $^{208}Tl$ activity of the $^{100}Mo$ source foils is equal to 68 $\mu$Bq/kg, at 90$\%$ C.L.. This limit, already better than the one measured by $\gamma$-ray spectroscopy using HP Ge detectors ($\leq \, 110 \, \mu$Bq/kg) is not so far from the limit required by the NEMO 3 radiopurity  criteria ($\leq \, 20 \, \mu Bq/kg$). We expect to reach the required limit within four months of data.

\begin{table}[htbp]
\begin{center}
\begin{tabular}{|l|c|c|c||}
\hline \hline
Channel  & Activities ($90\, \%$ C.L.)\\
\hline \hline
$e^-\gamma  $& $\leq\, 85 \, \mu$Bq/kg \\
 \hline
$e^-2\gamma$  & $\leq\, 78 \, \mu$Bq/kg \\
 \hline
$e^-3\gamma  $&$\leq\, 86 \, \mu$Bq/kg \\
 \hline

\hline

\end{tabular}
\end{center}
\caption{Upper limits ($90\, \%$ C.L.) on the reconstructed activities in $^{208}Tl$ of the $^{100}Mo$ source foils, using a sample of events representing 900 hours of acquisition.}
\label{resul_tl208_2}
\end{table}

\section{Double beta analysis}

Using the same sample of events, representing the 900 hours of our first test runs, we could perform preliminary analysis of the $\beta\beta2\nu$ and $\beta\beta0\nu$ signals. In this paper, this analysis is presented for $^{100}Mo$ only. Therefore, the events selected have to be emitted on the same point of a $^{100}Mo$ foil~: geometrical cuts on the vertex of the electrons emitted and temporal cuts (based on a $\chi^2$ analysis) devoted to the selection of electrons emitted inside the source foils, have been applied.

\subsection{Analysis of the $\beta\beta2\nu$}

\begin{figure}[h]  
         \centerline{\epsfxsize=4cm \epsfbox{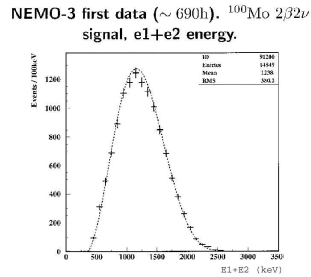}} 
         \caption{\label{bb2n_ene} Sum of the kinetic energies of the two electrons for the $\beta\beta2\nu$ events~: the dotted points represent the data and the line, Monte Carlo events}
\end{figure}

\begin{figure}[h]  
         \centerline{\epsfxsize=4cm \epsfbox{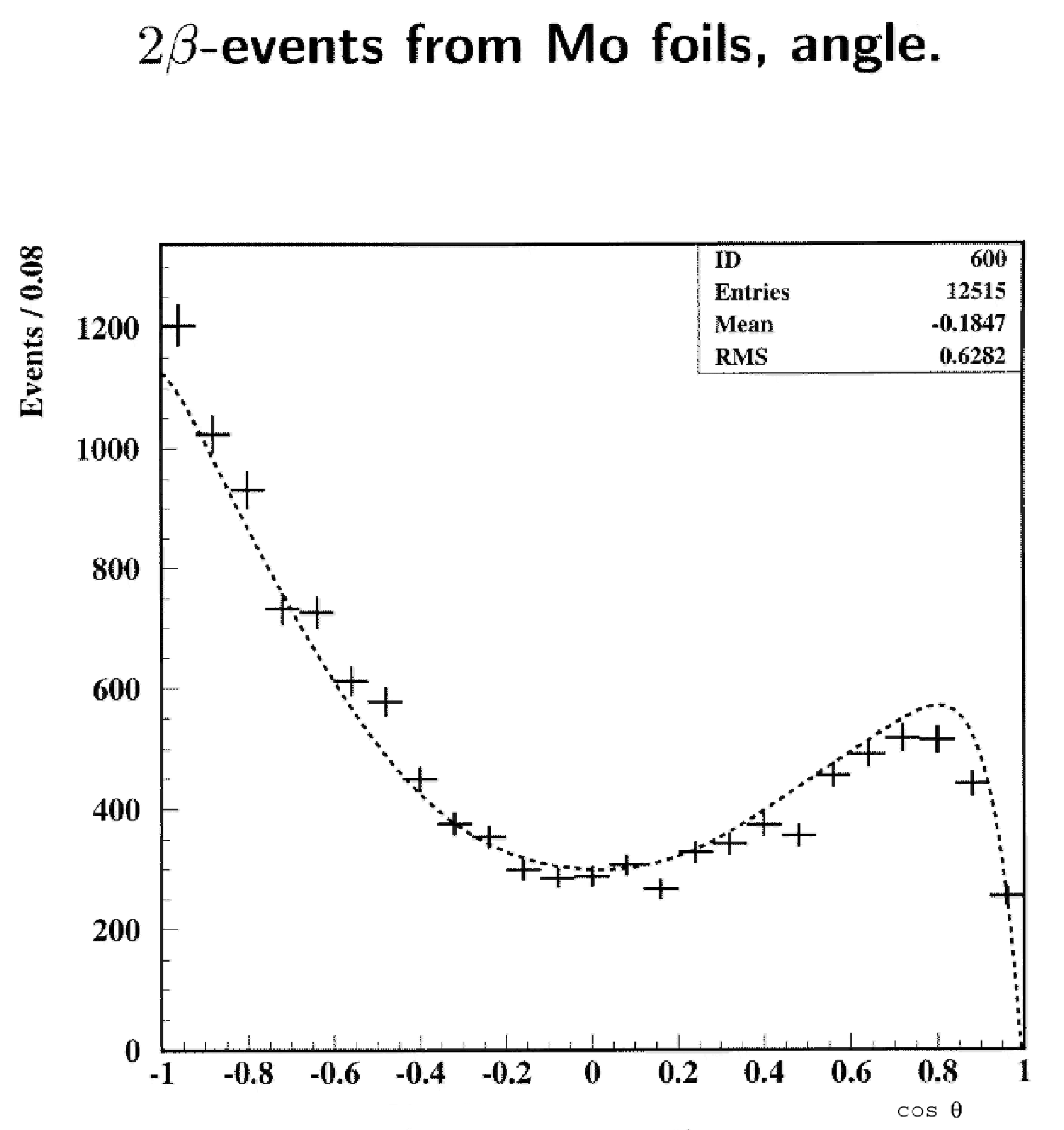}} 
         \caption{\label{bb2n_ang} Angular distribution for the $\beta\beta2\nu$ events~: the dotted points represent the data and the line, Monte Carlo events}
\end{figure}

Using the sample of events described below, representing 18000 events, we performed a preliminary analysis of $\beta\beta2\nu$ decay. The spectrum of the sum of the kinetic energies of both electrons is shown in Fig.\ref{bb2n_ene}, and the angular distribution is shown in Fig.\ref{bb2n_ang}~: there is a good agreement between data and Monte-Carlo. The half-life deduced from these data is equal to $8.\, \pm \, 0.08 (stat.)\, \pm\, 2 (syst.) 10^{18}$y~: it is already of the same order of magnitude than the results given by NEMO 2.

\subsection{Analysis of the $\beta\beta0\nu$}

Up to now, the analysis of the $\beta\beta0\nu$ signal by the NEMO collaboration was based on the study of the events in an energy window centered on the $Q_{\beta\beta}$ of the studied nucleus.  Another method, presented in this paper, is based on a maximum of likelihood method. The choice of this method has several strong motivations~: 

\begin{itemize}

\item The NEMO 3 detector is able to measure the three variables present in the differential cross section of the $\beta\beta$ processes~: the energies of the two electrons and the angle between the two tracks. Moreover, this detector reconstructs its own background. Therefore, a maximum of likelihood is the only way to take into account all this information.

\item We can constrain the tail of the spectrum of the $\beta\beta2\nu$. 

\item It is also a way to look at the coherence of the energy measurement, if we keep the $Q_{\beta\beta}$ as a free parameter of the likelihood.

\end{itemize}

\subsubsection{Principle}

Tridimensional probability distribution function have been built, using simulated events, for the different contributions to 2$e^-$ events~: the $\beta\beta0\nu$ signal, the $\beta\beta2\nu$, the $^{208}Tl$ and $^{214}$Bi, internal and external, background and the neutrons. The three variables are the sum of the kinetic energies of the two electrons - discriminating variable studied  above 2 MeV in this method -, the energy of the electron of minimal energy, and the angle between the two tracks. The likelihood is defined by the following expression, where X corresponds to one of the 7 contributions mentionned below, $x_X$ is the ratio of the number of 2$e^-$ events due to X to the total number of events $N_{tot}$, and $P^X_{3D}$, the tridimensional probability distribution function associated to X~:\\

\begin{center}
\fbox{${\cal L} = \,  \prod_{i\, =\,1}^{N_{tot}} (\sum_{X=1}^7 \, x_X P^X_{3D})$}
\end{center}

Tha amplitudes of the internal and external backgrounds, and of the neutrons, can be fixed since we can reconstruct them using other channels. Therefore, the maximisation of ${\cal L}$ gives us access to the contribution of the $\beta\beta0\nu$ signal, and to the value of the end-point of the studied nucleus ($Q_{\beta\beta}$).

\begin{figure}[h]  
         \centerline{\epsfxsize=4cm \epsfbox{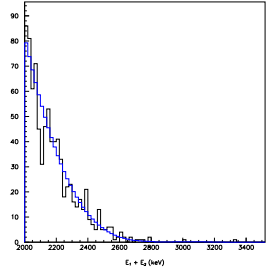}} 
         \caption{\label{resul} Sum of the kinetic energies of the two electrons for the $\beta\beta0\nu$ events~: the data and the fit given by the likelihood - in blue - are superposed}
\end{figure}

An illustration of the application of this method on the 900 hours of test-runs data is shown on Fig.\ref{resul}~: there is, again, a good agreement between data and the fit of them given by the analysis presented here. The sensitivity on the effective mass will be given as soon as we have a sufficient amount of data taken in stable conditions and with a neutron shield.

\section{Conclusion}

 With the first test runs taken by the NEMO 3 experiment, we have determined the main characteristics of the detector, in good agreement with what we expected~: the energy and time calibration have been performed and the performances of the tracking detector have been determined. Moreover, the preliminary analysis of the $^{208}Tl$ pollution of the $^{100}Mo$ source foils gave us an upper limit on this pollution, equal to $68\, \mu$Bq/kg , at $90\%$ C.L.~; the  limit required by NEMO 3 should be reached within four months of data. Finally, the $\beta\beta$ analysis show a good agreement between Monte Carlo and data, and will be applied to the data taken by the full detector, in stable conditions.

\end{document}